# A Comprehensive Study on the Applications of Artificial Intelligence for the Medical Diagnosis and Prognosis of Asthma

## Saksham Kukreja

*Student Researcher (Independent)*

research@sakshamkukreja.com

# 1 Research Abstract

An estimated 300 million people worldwide suffer from asthma, and this number is expected to increase to 400 million by 2025. Approximately 250,000 people die prematurely each year from asthma out of which, almost all deaths are avoidable. Most of these deaths occur because the patients are unaware of their asthmatic morbidity. If detected early, asthmatic mortality rate can be reduced by 78%, provided that the patients carry appropriate medication for the same and/or are in lose vicinity to medical equipment like nebulizers. This study focuses on the development and evaluation of algorithms to diagnose asthma through symptom intensive questionary, clinical data and medical reports. Machine Learning Algorithms like Back-propagation model, Context Sensitive Auto-Associative Memory Neural Network Model, C4.5 Algorithm, Bayesian Network and Particle Swarm Optimization have been employed for the diagnosis of asthma and later a comparison is made between their respective prospects. All algorithms received an accuracy of over 80%. However, the use of Auto Associative Memory Model (on a layered Artificial Neural Network) displayed much better results. It reached to an accuracy of over 90% and an inconclusive diagnosis rate of less than 1% when trained with adequate data. In the end, naïve mobile based applications were developed on Android and iOS that made use of the self-training auto associative memory model to achieve an accuracy of nearly 94.2%.

# 2 Introduction

Asthma is a chronic paroxysmal, often allergic disorder of respiration characterized by variable and recurring symptoms of breathlessness and wheezing, bronchospasm, and reversible airflow obstruction. It is one of the most important chest diseases, which is caused by contraction and inflammation of the airways in the respiratory system. It causes recurring periods of wheezing, chest tightness, shortness of breath, and coughing, due to obstruction of the lumen of a bronchiole by mucoid exudate, goblet cell metaplasia, and/or epithelial basement membrane thickening. Other common symptoms of asthma include insomnia (sleeplessness), daytime fatigue, reduced activity levels, and frequent school and work absenteeism.

According to the World Health Organization (WHO), the prevalence of asthma in different countries varies widely, but the disparity is narrowing due to rising prevalence in low and middle-income countries and plateauing prevalence in high-income countries. An estimated 300 million people worldwide suffer from asthma, with 250,000 annual deaths attributed to the disease. It is estimated that the number of people with asthma will grow by more than 100 million by 2025. If appropriate measures are not taken, the premature death toll is expected to increase by 125,000 due to asthma by 2025. Workplace conditions, such as exposure to fumes, gases or dust, are responsible for 11% of asthma cases worldwide. Occupational asthma contributes significantly to the global burden of asthma, since the condition accounts for approximately 15% of asthma amongst adults. It has also been observed that nearly 70% of asthmatic patients have allergies.

Asthma is a ubiquitous disease, which can be found in all the countries of this world. It is the most common chronic disease among children and its number of patients grow at a faster rate than any other major chronic disease due to an increase in pollution and population every day. Although it has a low mortality rate when compared to other chronic diseases, its large omnipresence poses a serious risk of sudden attacks, which can lead to death in just five minutes. If diagnosed early, patients can be prepared to handle abrupt breathless conditions with appropriate medication and awareness. The prognosis of the disease evaluates the complexity and risk of asthmatic morbidity, and equips the patient with appropriate knowledge and technology.

To diagnose a person's asthmatic morbidity early, novel Artificial Intelligence (particularly, Machine Learning) algorithms have been employed. Due to recurring updating of a disease characteristic database, these algorithms can also produce an automated prognosis, so that the user can understand the gravity of his/her morbidity. There are several benefits to using Machine Learning algorithms in the field of biomedicine. It eliminates the added dimension of mistakes committed due to human carelessness. The extensive amount of data available online also makes it possible for the algorithms to train themselves and achieve near perfect accuracy - something that is very difficult for any human to achieve. Furthermore, it is easy to replicate and can be shared globally, thus reducing costs associated with logistics and labor force.

For analysis of asthmatic morbidity, three different methodologies are combined to yield a meaningful result:

1. Symptom Intensive Questionary
2. Medical Reports
3. Chest Sound Analysis (like AdaBoost combined with Random Forest)

Although one of these methodologies is enough to diagnose bronchial disorders and future vulnerabilities, a combination is able to relate better to the disease database composed of past entries. The database when loaded with information that covers two or all three of these methodologies is better equipped to diagnose various bronchial disorders. In turn, the accuracy of the algorithms increases significantly.

For this study, algorithms like Backpropagation, Particle Swarm Optimization, Bayesian Network, C4.5 and Auto Associative Memory Model have been used. Later, a comparison is drawn to find the most suitable algorithm based on complexity, efficiency, accuracy, nonlinear data, explaining capacity, narrowing diagnostic possibilities and automatic assignment of probabilities.

**International Guidelines for Diagnosis of Asthma (EPR 3):**

As commissioned by the National Asthma Education and Prevention Program (NAEPP) Coordinating Committee, International guidelines recommend making asthmatic diagnosis based on typical asthmatic symptoms and identification of airway hyper-responsiveness (AHR) or variable airway obstruction, which prove to be asthma's key characteristics. Bronchial provocation test using nonspecific stimuli, such as methacholine or histamine, is useful for the determination of AHR. However, it is somewhat invasive in nature and is not easy to perform by general physicians. In addition, it is not available in primary healthcare facilities or even in many general hospitals. Bronchodilator response (BDR) to short acting β-2-agonists is a valuable test to evaluate variable airway obstruction, which is only useful in patients with reduced lung function at the time of visit. Without information about AHR and BDR of a patient, a physician has to make a diagnosis of asthma based on respiratory symptoms and physical examination only. Thus, the decision is not dependent only on the objective evidence of asthma but also on the experience of the doctor.

**Outline:** The remainder of this paper is organized as follows. Section 3 contains the symptom intensive questionary designed for preliminary diagnosis. Section 4 describes how medical reports are analyzed and used by the ANN. Section 5 describes the machine learning algorithms that are put to use, their design, implementation and comparison. Section 6 presents the results achieved, followed by Section 7 that describes the implementation of the research via mobile apps. This is followed by Conclusion, Acknowledgements and References in Sections 8, 9 and 10 respectively.

# 3  Symptom Intensive Questionary

The American Academy of Allergy, Asthma, and Immunology (AAAAI) estimated that nearly two-third of asthmatic patients in the world are unaware of their asthmatic morbidity. The principal factor that is responsible for this statistic is misdiagnosis of asthma as common cold, pneumonia, acute bronchitis, chronic obstructive pulmonary disease (COPD) etc. Hence, an exhaustive questionary is created for the user that utilizes data from the disease characteristic database for distinct diagnosis and prognosis of asthma.

The user based questionary has been developed as follows:
Questions have been divided into groups and each group has been assigned a priority factor. The priority factor functions as the group's score. Let's consider a case where there are $n$ groups. Let the priority factor of each group be $a_i$ where $i \in [1 \ldots n]$. Let there be $m_i$ questions in the $i^{th}$ group where $i \in [1 \ldots n]$. Let the response of the $j^{th}$ question of the $i^{th}$ group, $y_{ij}$ be stored as 1 (for yes) and 0 (for no) in elements $e_{ij}$ where $i \in [1 \ldots n]$ and $j \in [1 \ldots m_i]$.

Based on the user responses, Asthmatic Questionary Computable Score ($A$) is calculated as follows:

$$A = \sum_{i=1}^{n} \left( a_i \cdot \sum_{j=1}^{m_i} e_{ij} \right)$$

$A$ lies in the range $[0 \ldots x]$ where $x = \sum_{i=1}^{n}(a_i \cdot m_i)$.

A binary response matrix, $M$ with $x$ elements is also created which stores responses in the following manner:
The response of $y_{pq}^{th}$ question, $e_{pq}$ is stored in $a_p$ consecutive elements indexed from $[h + 1 \ldots h + a_p]$ where $h = \sum_{i=1}^{p-1}(a_i \cdot m_i) + a_p \cdot (q - 1)$.

Using the above matrix, $A$ can also be calculated as $A = \sum_{i=0}^{x} t_i$ where $t_i$ is the element of the matrix $M$ stored in the $i^{th}$ index where $i \in [1 \ldots x]$.

After comprehensive research on symptoms, causes and effects of asthma, and with the suggestions of doctors who have mastered respiratory medicine, the following questionary was developed:

| Group | Symptom Intensive Questions | Priority Factor |
|---|---|---|
| I | A. Can a distinctive wheezing sound be heard after you run/exercise?<br>B. Do you suffer from any sort of allergies?<br>C. Do you breathe with your mouth more often than your nose?<br>D. Do you feel random sudden pains in your chest that make you feel suffocated?<br>E. Do you take more than 30 breaths in a minute?<br>F. **The Interchange Test:**<br>Cover your left nostril and inhale as much air as you can. A slight pain in the chest is normal. Hold your breath for 15 seconds. Now, cover your right nostril and release air from the left nostril as quickly as you can. Breathe normally and relax for one minute. Tap on yes if one or more of the following is true:<br>i) You feel suffocated.<br>ii) You feel a pain in your chest.<br>iii) You had to cough after releasing the air.<br>iv) You hear wheezing or feel very anxious. | 6 |
| II | G. Has there been any incident when you had to wake up in the middle of the night due to excessive coughing?<br>H. Do you feel that you are out of breath episodically (like every winter, every night etc.)?<br>I. Do you have swelling in your feet, ankles and/or hands?<br>J. Do you see blood when you spit or in you sputum after you cough?<br>K. Do you have to put in effort to breathe? | 5 |
| III | L. Do you face any problems in breathing when you are suffering from fever but not cold?<br>M. Do you smoke?<br>N. Do you work in any smoke/ceramic/stone/cement factory/industry?<br>O. Does cough produce any sputum?<br>P. Do you feel heat/burning sensation in your chest?<br>Q. Do you feel obstruction during breathing due to mucus? | 4 |
| IV | R. Do you have anxiety issues?<br>S. Do you have problems in sleeping? (E.g. Unable to sleep or wake up early, insomniac feeling etc.)<br>T. Do you experience some coarseness in your throat after you wake up? | 3 |
| V | U. Do you feel tightness/ restriction in your chest while/after eating a full meal?<br>V. Do you live in a highly polluted city? | 2 |
| VI | W. Do you get squeaky when you just start speaking?<br>X. Do you feel pain in the chest after heavy exercises? | 1 |

The aforementioned questionary is characterized by the following attributes:

$n = 6$;

$a_1 = 6$, $a_2 = 5$, $a_3 = 4$,

$a_4 = 3$, $a_5 = 2$, $a_6 = 1$;

$m_1 = 6$, $m_2 = 5$, $m_3 = 6$,

$m_4 = 3$, $m_5 = 2$, $m_6 = 2$;

$\therefore x = \sum_{i=1}^{n}(a_i . m_i) = 100$

Asthma is usually predicted by the doctor only on the basis of a person's response to a subset of this questionary. As stated earlier, the diagnosis decision is not dependent only on the objective evidence of asthma but also on the experience of the doctor. If this questionary were utilized in diagnosis without any algorithms, the results obtained would be equivalent to that of an unexperienced doctor. On testing without machine learning algorithms, the questionary yielded an accuracy in the range of $[50\% \ldots 60\%]$. This was done by using a threshold, $\phi$ on the Asthmatic Questionary Computable Score $(A)$. However, due to its poor results, it cannot be used as the sole predictor of a person's asthmatic morbidity. Therefore, artificially intelligent system must be trained with adequate training sets to yield results equivalent to that of an experienced, professional doctor.

In a small survey of nearly 50 doctors (general physicians), we observed that during diagnosis checkups, it was difficult to confirm asthmatic morbidity with full confidence. There is always a chance that he/she misdiagnoses asthma to be another respiratory disorder. Therefore, to supplement the original questionary, an optional professional questionary is used to increase the accuracy of algorithms.

| Symptom Intensive Questions | Priority Factor |
|---|---|
| 1. Have you had wheezing associated with dyspnea?<br>If yes, is it provoked by:<br>a) Cold Air<br>b) Smoke or Air Pollution<br>c) Exercise<br>d) Upper Respiratory Infection<br>e) Nocturnal Aggravation<br>f) Concurrently with coughing. | 8<br><br>4<br>4<br>3<br>3<br>3<br>2 |
| 2. Have you had paroxysmal coughing? | 5 |
| 3. Have you had dyspnea without wheezing? | 5 |
| 4. Have you had wheezing without dyspnea? | 5 |
| 5. Have you had fluctuation of exacerbation and improvement? | 8 |

The score B and a matrix N are determined through this questionary. Although most of these questions have been covered before, the above questionary makes sure that the person identified his problem properly. Therefore, the accuracy is improved.

It is mainly the matrices that are passed in as features to the Artificial Neural Network and other related algorithms. Some secondary features, like bronchial obstruction subscore, pollutant effect subscore etc., which are calculated by the summation of elements of certain predefined indices, are also passed. The calculations involved are done just after matrix creation and hence, the values (subscores) obtained are passed as input features.

# 4 Medical Report Analysis

## 4.1 Pulmonary Function Tests (or Lung Function Tests)

The pulmonary function testing involves a complete evaluation of the patient's respiratory system. The primary purpose of pulmonary function testing is to identify the severity of pulmonary impairment. Pulmonary function testing has diagnostic and therapeutic roles and helps clinicians answer some general questions about patients with respiratory disorders, however, the test results alone cannot identify whether or not a person is suffering from a respiratory disease. They can only provide certain pointers to the diagnostician regarding a person's respiratory morbidity. Moreover, it is still the doctor's job to find the particular respiratory disease from which his patient is suffering. Even after analyzing medical reports, there is a significant chance of misdiagnosis of asthma as Chronic Obstructive Pulmonary Disease, Chronic Shortness of Breath, Restrictive Lung Disease, Impairment, Bronchitis, Pneumonia, or Tuberculosis. Therefore, the fact that "the decision is not dependent only on the objective evidence of asthma but also on the experience of the doctor" still holds true. In this case too, machine intelligent algorithms need to be employed for proper diagnosis and prognosis. The statistics act as features and are fed as inputs to ANN.

The pulmonary function tests involve the following:

1. **Spirometry:** Spirometry includes tests of pulmonary mechanics – measurements of FVC (Forced vital capacity: the determination of the vital capacity from a maximally forced expiratory effort), FEV1 (Volume that has been exhaled at the end of the first second of forced expiration), FEF (Forced expiratory flow related to some portion of the FVC curve; modifiers refer to amount of FVC already exhaled) values, FIFs (Forced inspiratory flow: the volume inspired from RV at the point of measurement.), and MVV(Maximal voluntary ventilation: volume of air expired in a specified period during repetitive maximal effort). Measuring pulmonary mechanics assesses the ability of the lungs to move large volumes of air quickly through the airways to identify airway obstruction. The measurements taken by the spirometry device are used to generate a pneumotachograph that can help to assess lung conditions such as asthma, pulmonary fibrosis, cystic fibrosis, and chronic obstructive pulmonary disease.

2. **Plethysmography test:** Plethysmography is a lung performed to measure the compliance of the lungs by determining how much air the lungs can hold. For this test, the person sits or stands in a small booth and breathes into a mouthpiece. The lung volume is recorded by measuring the pressure in the booth.

3. **Airways Resistance:** It describes the resistance of the respiratory tract to airflow during inspiration and expiration.

## 4.2 CT Scan

Respiratory disorders can be sometimes identified via Chest Computer Topography images. This usually occurs due to two reasons:

1. **Visible Cause:** The cause of the disease is an explicitly visible factor that has been recorded in the CT image. Eg. Tumor, Muscular Stiffness etc.
2. **Visible Effect:** The disease resulted in a formation (like an air-entrapment cage) that is visible in the CT image.

In order to separate a Region of Interest (ROI), Support Vector Machine was used along with image processing and image analysis.

Threshold used: $T_{i+1} = \frac{1}{2}(u_a + u_b)$

The process involved is as follows:

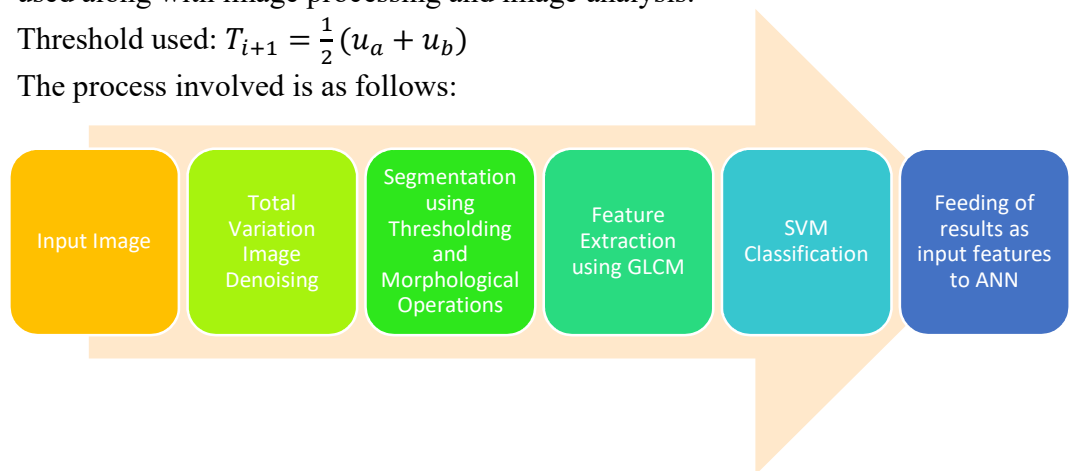

The features passed into the ANN and its algorithms are calculated as follows:

| Feature | Formula | Function |
|---|---|---|
| Area (A) | - | Total no. of pixels in ROI. |
| Convex Area (CA) | - | Total no. of pixels in the convex region of ROI. |
| Equivalent Diameter (ED) | $\sqrt{\frac{4A}{\pi}}$ | Diameter of circle with same area as that of ROI. |
| Solidity (S) | $\frac{A}{CA}$ | Ratio of Area to Convex Area. |
| Energy (E) | $\sum_{k=0}^{n} p^2(i,j)$ | Summation of squared elements in the GLCM. |
| Contrast (C) | $\sum_{i=0}^{N}\sum_{j=0}^{N}((i-j)^2 p(i,j))$ | Contrast between an intensity of pixel and its neighboring pixels over the whole ROI. |
| Homogeneity (H) | $\sum_{(i,j)} \frac{p(i,j)}{1-|i-j|}$ | Closeness of the distribution of elements in the GLCM to the GLCM of each ROI. |
| Eccentricity | - | Ratio of the distance between the foci of the ellipse and its major axis length. |

## 4.3 Impulse Oscillometry

Impulse oscillometry is a noninvasive and rapid technique requiring only passive cooperation by the patient. Pressure oscillations are applied at the mouth to measure pulmonary resistance and reactance. Pressure resistance and pressure reactance are passed as input features to the ANN and its related algorithms.

Feeding features from medical reports is optional for the user. However, if the user decides on entering the values, the accuracy increases to nearly 96.9% if all values are entered. In the mobile application, it is possible to add some selected elements from the medical reports section in order to accommodate the test changes and make maximum use of the available information. This is because the inputs created by medical analyses into the matrix are preceded with a Boolean variable. If the user enters a value, the variable is turned to true (1) and the test result is fed to the matrix. Therefore, the final input will also depend on the number of true Boolean literals and it changes the course of further algorithms. Mostly, missing on the CT Scan does not degrade the accuracy of asthma detection. However, providing those results can help the system analyze whether or not a person is having lung cancer in addition to facilitating it with more information for asthmatic diagnosis.

# 5 Use of Machine Intelligent Algorithms

## 5.1 Development of ANN

The ANN employs MLNN and PNN methods as they prove to have enhanced probability of predicting respiratory diseases.

A neural network is created using the in-built toolbox in MATLAB. It is further exported to Python for use in the mobile application. All the features that are passed to the ANN are sent as inputs to the input layer. A more resolved input is created using techniques like CSAAM, which are mentioned below.

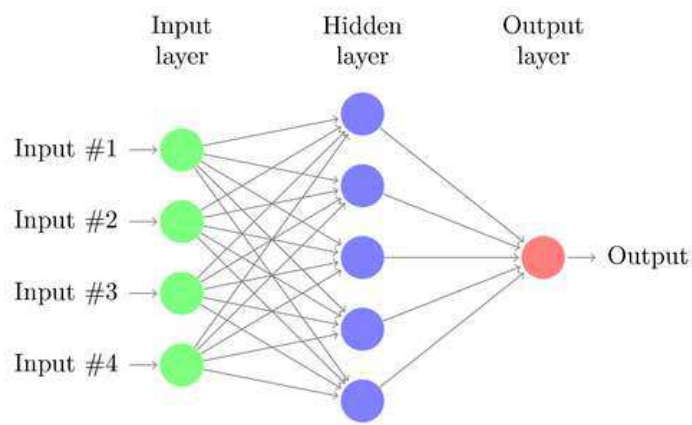

In order to provide an operational input encompassing all identifying factors and features that are passed into the artificial neural network, a $m \times n$ matrix is set up where each element acts both as an information storage and a referential node. The value of $m$ and $n$ is determined based on the course followed during the execution of the set-up.

This matrix, which reflects all the input features that are to be passed into the Artificial Neural Network and its related algorithms, is termed as the reflective input matrix.

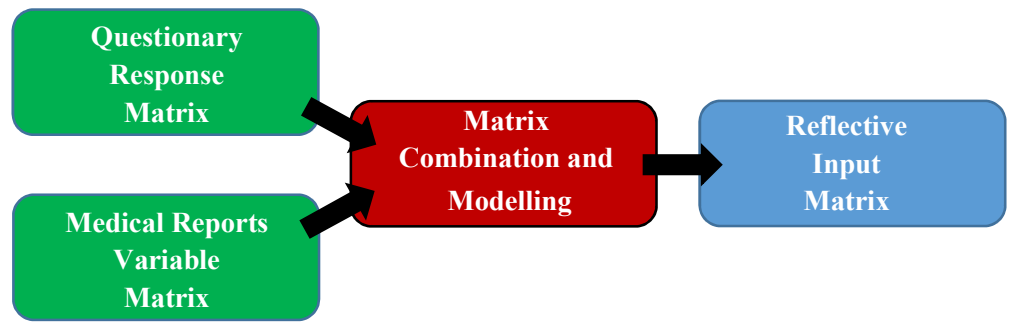

As Artificial Neural Networks (ANNs) constitute a major class of medical computing, their utilization for diagnosis of asthma will allow experienced professionals in medical centers to easily adapt to the new technology.

To check the results of the algorithms, the ANN was converted to a standalone service and deployed on the web. For all relevant tests, training sets and testing sets were used in a fixed ratio and the results were recorded. The results were compared based on the following values:

| Characteristic | Formula | Description |
|---|---|---|
| **True Positive (TP)** | - | Number of positive cases correctly diagnosed. |
| **True Negative (TN)** | - | Number of negative cases correctly diagnosed. |
| **False Positive (FP)** | - | Number of negative cases incorrectly diagnosed as positive. |
| **False Negative (FN)** | - | Number of positive cases incorrectly diagnosed as negative. |
| **Inconclusive** | - | Number of cases that could not be diagnosed. |
| **Sensitivity** | $\dfrac{TP}{TP + FN}$ | It measures the proportion of positives that are correctly identified as such. |
| **Specificity** | $\dfrac{TN}{TN + FP}$ | It measures the proportion of negatives that are correctly identified as such. |
| **Positive Prediction Rate (PPR)** | $\dfrac{TP}{TP + FP}$ | It is the ratio of correctly diagnosed positives and total diagnosed positives. |
| **Negative Prediction Rate (NPR)** | $\dfrac{TN}{TN + FN}$ | It is the ratio of correctly diagnosed negatives and total diagnosed negatives. |
| **Matthew's Correlation Coefficient (MCC)** | | MCC is correlation coefficient between the observed and predicted binary classifications; it returns a value between −1 and +1. It is calculated by the formula: $$\dfrac{TP \times TN - FP \times FN}{\sqrt{(TP + FP)(TN + FN)(TN + FP)(TP + FN)}}$$ |
| **Accuracy** | | It is the ration of correctly diagnosed cases and total number of cases. It is calculated by the formula: $$\dfrac{TP + TN}{TP + FP + TN + FN}$$ |
| **F₁ Score** | | It is another characteristic to measure the efficiency (accuracy). Its value is equal to the harmonic mean of PPR and Sensitivity. It is calculated by the formula: $$2 \cdot \dfrac{PPV \cdot TPR}{PPV + TPR} = \dfrac{2TP}{2TP + FP + FN}$$ |

## 5.2 Context Dependent Auto-Associative Memory

In order to use a disease database adeptly and individually match reflective input matrices with the elements of the database, a model called the context dependent (auto) associative memory is employed. Its ease of implementing basic logical operations of propositional calculus and computing fundamental operations of modal logic add to the merits of efficiently modulating the matrix correlation memories, which it already possesses. In this model, vectorial content and the key vector are combined to conform to a Kronecker Product.

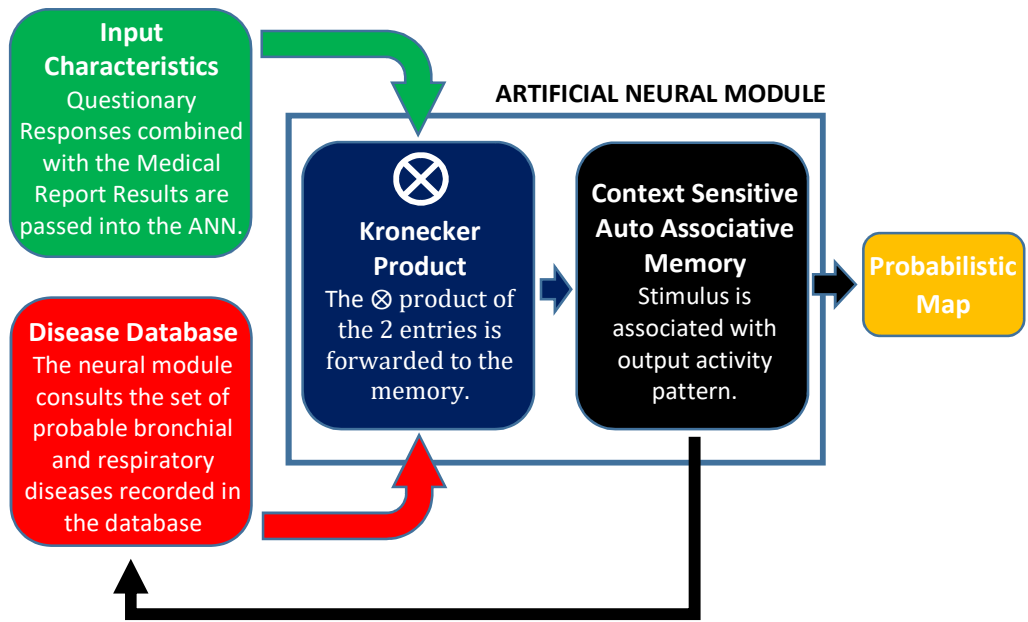

It is evident that this algorithm is very well suited to the medical industry, especially in diagnosis and prognosis as it computes the output after relating it with a large database and then adds a related output to the preexistent disease database. The best way to represent the inputs is through vectors and/or matrices. The vectors are associated with the new info added by the user and the set of possible prognoses depending on the previously loaded database respectively.

The vectors are passed into the artificial neural module, which consists of two parts:-
1. Kronecker Product ($\otimes$), which establishes a proper relationship with the database by linking all possibilities into a single input to the associative memory.
   It is defined as $A \otimes B = a_{ij} \times B$, where each element of $A$ is multiplied by the entire matrix $B$. Therefore, for $A_{m \times n}$ and $B_{e \times f}$, the resultant matrix is of the order $me \times nf$.

2. Context Dependent Associative Memory ($\psi$), which narrows the set of all possible options to the most relevant ones. The linear combination is processed with mathematical functions to reach to a conclusive result. It acts as follows:

$$\psi = \sum_{i=1}^{k} t_i \left( t_i \otimes \sum_{j(i)} s_j \right)^T$$

Where $t_i$ are column vectors mapping $k$ different diseases, (the set $\{t\}$ is chosen to be orthonormal), and $s_j(i)$ are column vectors mapping signs or symptoms accompanying the $i^{th}$ disease (also an orthonormal set). The sets of symptoms corresponding to each disease can overlap.

## 5.3 $C4.5$ Algorithm with Bayesian Networks

$C4.5$ Algorithm is one of the best algorithms that can be employed in AI Medical Diagnosis due to its basis of decision trees. It uses the concept of information entropy. The training data is a set of already classified samples. It may be denoted by $\Omega = \omega_1, \omega_2, \omega_3, \omega_4, \cdots$, where each sample is a vector $\omega_i = \lambda_1, \lambda_2 \cdots$. Here, $\lambda_1, \lambda_2 \cdots$ represent the attributes of the sample. The set $\{\Omega\}$ is augmented with the vector $\Gamma = \tau_1, \tau_2 \cdots$ where $\tau_1, \tau_2 \cdots$ represent the class to which each sample belongs. $C4.5$ is a statistical based decision tree algorithm that employs divide and conquer approach to grow decision trees. It was formed by pruning of $ID3$ algorithm to reduce noise as well as compute continuous data.

When $C4.5$ algorithm is coupled with Bayesian Networks, which is a purely probability based algorithm, it also computes the noise level and helps in matching the exact disease from the inputs provided. The datasets that are used in this method can be represented by using multiple file formats:-

1. **Asthma.set:** This is primarily used as the disease database and will contain the classification of the disease. The parameters - including symptoms and medical data in discrete or continuous form - are supplied and updated with continuous use.
2. **Asthma.data:** This is used as the training set for the $C4.5$ + BN model.
3. **Asthma.test:** This represents the test data.

PAGE 15 OF 27

## 5.4 Backpropagation Model

Backpropagation is a well-known training algorithm for multi-layer neural networks. It defines rules of propagating the network error back from network output to network input units and adjusting network weights along with this back propagation. It requires lower memory resources than most learning algorithms and usually gets acceptable results, although it can be too slow to reach the error minimum and sometimes finds not the best solution.

Back-propagation can also be considered as a generalization of the delta rule for non-linear activation functions and multilayer networks. There are a variety of backpropagation models; among them we considered Incremental backpropagation method to implement the expert system. This is a variation of the Back Propagation where the network weights are updated after presenting each case from the training set, rather than once per iteration. This is originally invented variant of back propagation and sometimes referred to as Standard Back Propagation. It can be the most preferred algorithm for large data sets.

## 5.5 Particle Swarm Optimization

Particle swarm optimization (PSO) is a population based stochastic optimization technique developed by Dr. Eberhart and Dr. Kennedy in 1995, inspired by social behavior of bird flocking or fish schooling. PSO shares many similarities with evolutionary computation techniques such as Genetic Algorithms (GA). The system is initialized with a population of random solutions and searches for optima by updating generations. However, unlike GA, PSO has no evolution operators such as crossover and mutation. In PSO, the potential solutions, called particles, fly through the problem space by following the current optimum particles.

PSO is initialized with a group of random particles (solutions) and then searches for optima by updating generations. In each iteration, each particle is updated by following two "best" values. The first one is the best solution (fitness) it has achieved so far. This value is called *pbest*. Another "best" value that is tracked by the particle swarm optimizer is the best value, obtained so far by any particle in the population. This best value is a global best and called *gbest*. When a particle takes part of the population as its topological neighbors, the best value is a local best and is called *lbest*.

## 5.6 Comparative Analysis of Intelligent Algorithms

The aforementioned models were used in the development of artificially intelligent systems and appropriate inputs were provided to test their performance. The models were then refined for the systems to diagnose asthmatic morbidity with the consideration of the symptoms and clinical data entered into the artificial neural network as input parameters. The reports of the patients were analyzed, their questionary responses were recorded and the results were derived from various algorithms. The results derived can be condensed into the following table:

| Algorithm | Accuracy | Complexity | Efficiency | Accuracy | Data analysis | Probability Assigning |
|---|---|---|---|---|---|---|
| CDAAM | $86 \pm 3\%$ | High | V. Good | Excellent | Good | Good |
| C4.5 (+BN) | $83 \pm 2\%$ | Moderate | V. Good | Good | Good | Normal |
| BP | $81 \pm 2\%$ | Moderate | Low | Normal | Moderate | Normal |
| PSO | $84 \pm 4\%$ | Optimal | Good | Good | Good | Good |

It is evident that the Context Dependent Auto Associative Memory model is the perfect model for diagnosis when considered on grounds of validity, reliability, effectiveness, and accuracy of outcomes that were mapped with the knowledge of experts. Therefore, the CDAAM was the major system to be deployed for this study. It has been extensively used to derive results in the rest of the paper.

# 6 Results

Based on the above comparison, it is clear that Auto-Associative Memory, which is underpinned by the matrix correlation technique, is the most successful and the most promising method for the diagnosis and prognosis of asthma. Therefore, it was extensively tested on the standalone diagnostic server and the results of all relevant implementations were recorded. The setup was tested majorly in two situations: one with only the symptom intensive questionary, and the other with both the symptom intensive questionary and the medical reports. A breakup of 60 ∶ 40 and 50 ∶ 50 was used for training and testing of the algorithm without medical reports and with medical reports respectively. The data presented here is from the testing period of the algorithm. The results obtained were as follows:

I.  **Without Medical Reports:**

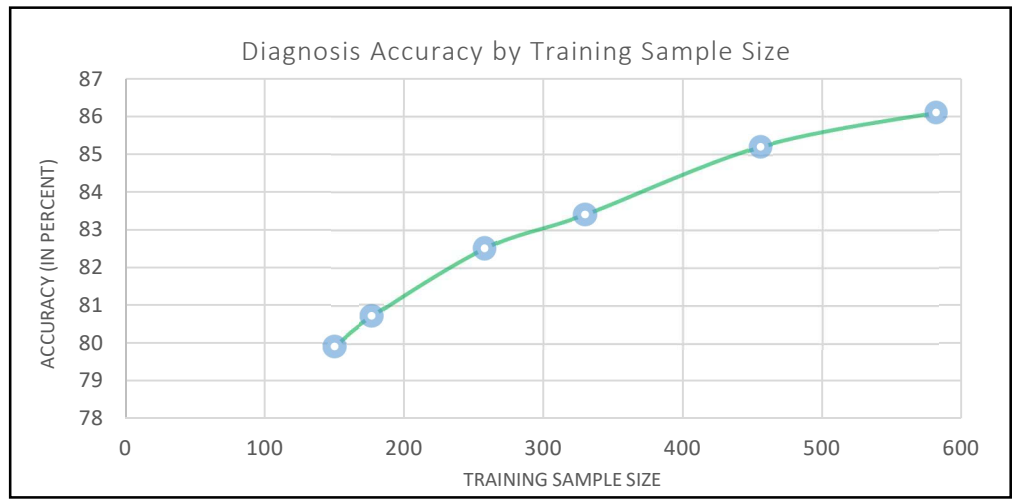

Even without medical reports, the setup delivered excellent accuracy that increased substantially with increase in the Training Sample Size. It reached to an accuracy of 86.1%.

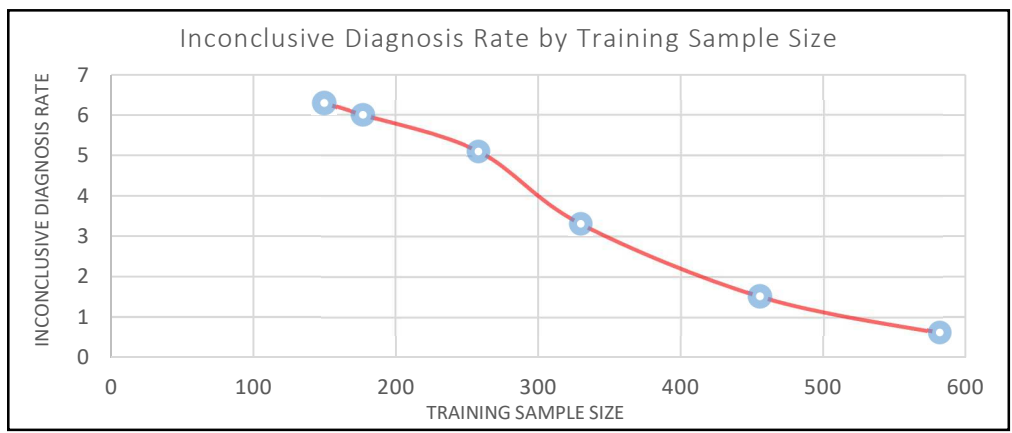

The decrease in the inconclusive diagnoses with increase in training sample size is also remarkable. When trained with a sample size of nearly 500, the inconclusive diagnosis rate was observed to be below 1%.

The performance of the setup is summarized as follows:

| Overall Outcome | Overall Results | | | | |
|---|---|---|---|---|---|
| | Positive | | Negative | | Sensitivity: 92.24% |
| | TP = 321 | FP = 42 | TN = 213 | FN = 27 | |
| | | | | | Specificity: 83.53% |
| | Inconclusive = 3 | | | | |
| | | | | | Matthew's Correlation Coefficient: |
| | Positive Prediction Rate: 88.42% | | Negative Prediction Rate: 88.75% | | 0.7647 |

The **Accuracy** of the setup was calculated to be **89%**.
The **F$_1$ Score** of the setup was calculated to be **0.9029** or **90.29%.**

II. **With Medical Reports**

Although the setup opts for different paths based on the number and choice of medical values reported, the following analysis was done with maximum possible input. CT Scan was intentionally ignored in 50% of the cases to study its effects on the algorithm's accuracy. The following results were obtained:

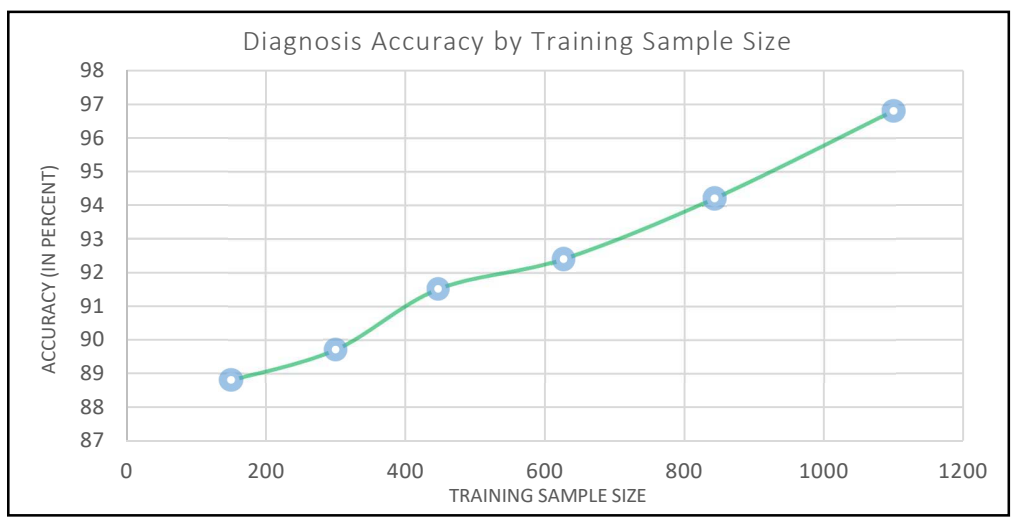

The above graph clearly points to the near-perfect accuracy that can be achieved on training the algorithm with a large training set. Presently, the setup reached to an accuracy of 96.81%.

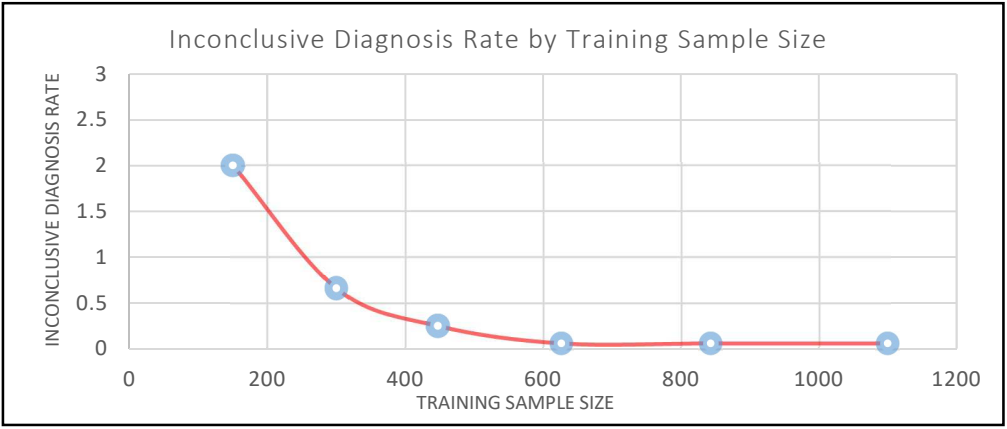

With increase in matrix features and training sample size, the inconclusive diagnosis rate dropped to zero. Therefore, inputs with both the questionary and medical reports account for an increased accuracy by 9-10% and a 100% result display in comparison to the inputs without medical reports. To establish relation with CT Scans, the accuracy was improved by 0.1-1.5% in total. However, CT Scans proved to be crucial in diagnosis of lung tumor, which is often misdiagnosed as COPD or asthma.

The performance of the setup is summarized as follows:

| Overall Outcome | Overall Results | | | | |
| --- | --- | --- | --- | --- | --- |
| | Positive | | Negative | | Sensitivity: 97.69% |
| | TP = 636 | FP = 20 | TN = 429 | FN = 15 | |
| | | | | | Specificity: 95.54% |
| | Inconclusive = 0 | | | | |
| | | | | | Matthew's Correlation Coefficient: |
| | Positive Prediction Rate: 96.95% | | Negative Prediction Rate: 96.62% | | 0.9341 |

The **Accuracy** of the setup was calculated to be **96.81%**.
The **F$_1$ Score** of the setup was calculated to be **0.9732** or **97.32%.**

# 7    Mobile Applications (Android and iOS)

To facilitate accessibility of this process to all the people across the globe, the ANN was converted to a standalone service and deployed on the web. This service was then used to initiate the diagnostic procedures from the inputs entered into the application. The applications are being developed both on Android and on iOS to reach a wider audience. They are kept under 5 MB so that they can be space efficient as well. In the future, we plan to introduce support for multiple languages to ensure that this technology penetrates through the linguistic barriers and serves its purpose of helping all the people in need.

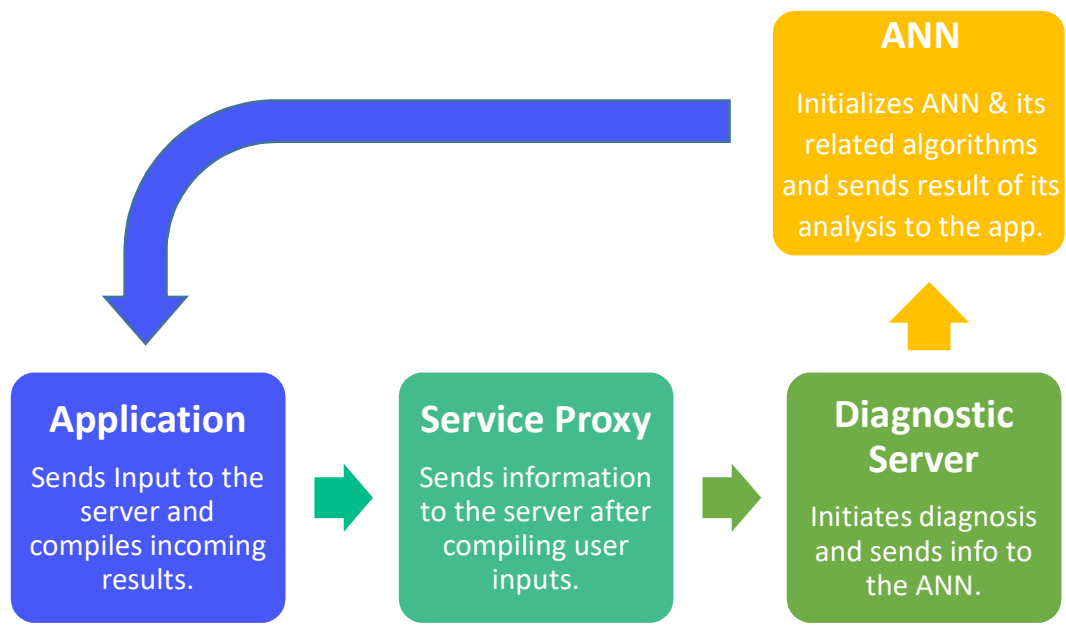

These applications allow the user to tap yes or no for every question in the questionary. Post answering, they allow the user to either enter the report results himself/herself or forward a response sheet to his doctor, who can fill it up on the same application. The CT Scan can be uploaded after filling in the medical reports. The results are computed and is displayed to the user. The user can also forward the result to his doctor to update him with the computed diagnosis and prognosis to allow efficient medical care in the future. The app also allows the user to rate its accuracy so that the dataset can be used for further training of the algorithms.

Currently, it is being developed and it would soon be published on App Store, Google Play and Amazon.

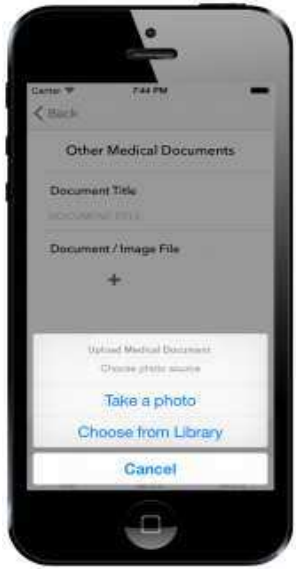

**A page from the iOS application for asthmatic diagnosis and prognosis:**

After the user inputs all the responses of the questionary, one is directed to a page where he/she may upload the results of the medical reports, CAT scans etc. to facilitate the application with more data and get the most accurate result possible. After completing the input sections, when the user taps on the **Get Result** button, the user will be redirected to a page where he/she will be able to see his/her own report. The user then has the option to forward this report to his doctor, if at all required.

# 8 Conclusion

The algorithms achieved a high accuracy: The ANN, when deployed with CDAAM and used with medical reports, achieved an extremely high accuracy of 96.81%. Even when the artificial neural network was used without medical reports, the algorithms managed to reach an accuracy of 89%. These statistics clearly point towards the ease of implementation offered by this innovation. The algorithms also succeeded in achieving a very low false positive and false negative rate, indicating that the experiments were a success. The ever-learning nature of the algorithms makes it possible for them to achieve near perfect accuracy by increasing their training sizes - something that a proposed online network facilitates.

The system implies the previous coding of a set of diseases and its corresponding semi-logic findings in individual basis of orthogonal vectors. The model presented in this communication is only a minimal module able to evaluate the probabilities of asthmatic diagnoses when a set of signs, symptoms and reports is presented to it.

The experiment also proved that CDAAM was the most promising algorithm for medical diagnosis in comparison to $C4.5$, Backpropagation, Particle Swarm Optimization etc. At the same time, the statistics reflect that PSO can be an extremely important alternative to the same as it ended with close results and optimal complexity. Particle Swarm Optimization could be altered in some respects so as to increase its accuracy and efficiency in the diagnosis of other diseases that may require more complicated referencing and networking.

A clear advantage of this system is that the probability assignment to the different diagnostic possibilities in any particular clinical situation does not have to be arbitrarily assigned by the specialist, but is automatically provided by the system, in agreement with the acquired experience.

Although previous work has succeeded in building computer aided systems for diagnosis (CAD) of diseases, its implementation in asthmatic diagnosis has only recently gained traction. The results achieved in this study are comparable to the seminal works in this field. In addition to the well-performing algorithms, the exhaustive questionary and deployment of an artificially intelligent server, the construction of a naive mobile application in the case of asthmatic diagnosis and prognosis is something that people will greatly benefit from. The mobile application allows people directly access and take diagnostic tests on their mobile phones, hence eliminating the costs associated with logistics and consultation. This is a boon in a world where consultation fees are sky-rocketing. In addition to this, a mobile application makes it feasible to carry out reliable tests in resources poor areas where taking exhaustive tests is not at all possible.

The scope of this experiment can be taken further into the diagnosis of other medical diseases, which are ubiquitous and ambiguous with respect to diagnosis.

I conclude that CDAAM model is a promising alternative in the development of accuracy diagnostic tools. I expect that its easy implementation stimulates groups of medical informatics to develop this model at real scale than other machine learning algorithms.

# 9 Acknowledgements

I would like to express my deepest gratitude to all the people who have helped me in this research project. First of all, I would like to thank Prof. Kentaro Shimizu and Prof. Takano for mentoring me in this project and facilitating me with all the resources that I needed at The University of Tokyo. Their support was the quintessential factor that underpinned the success of this project. I would also like to thank Dr. Takahide Nagase for letting me use the UTokyo Hospital Database Statistics, something without which, meaningful results could not have been obtained. Further, he assisted me in finding other repositories which could facilitate me with relevant statistics.

Apart from the people who helped me directly in the academic part of this project, I would like to extend my thanks to my family and friends, who sustained my belief in my work. Finally, I would also like to thank Prof. Andrew Ng, whose online course on Machine Learning opened the door to Artificial Intelligence for me.